\documentclass[a4paper]{article}
\usepackage{amsmath,amssymb,theorem,fullpage}
\usepackage[shortcuts]{extdash}
\usepackage[unicode]{hyperref}

\usepackage{csquotes}
\usepackage[backend=biber,maxbibnames=99,sorting=none]{biblatex}
\bibliography{cymoduli.bib}

\usepackage{color}
\usepackage{lipsum}
\usepackage{amssymb,amsmath,amsfonts,epic,eepic,float}

\usepackage[mathscr]{eucal}

\usepackage{rotating,epsfig,indentfirst,array,varioref}
\usepackage{appendix,marginnote,tikz,pgf,mathtools}
\usepackage{setspace}

\usepackage{authblk}

\makeatletter
\def\@@bfil{\leaders \vrule \@height \ht\z@ \@depth \z@ \hfill}
\def\@bLfil{\@@bfil}
\def\@bRfil{\@@bfil}
\def\resetbraceratio{\gdef\@bLfil{\@@bfil}\gdef\@bRfil{\@@bfil}}
\def\setbraceratio#1#2{
  \let\@bLfil\relax
  \multido{\iA=1+1}{#1}{\gappto\@bLfil{\@@bfil}}
  \let\@bRfil\relax
  \multido{\iA=1+1}{#2}{\gappto\@bRfil{\@@bfil}}
}
\def\upbracefill{$\m@th\setbox\z@\hbox{$\braceld$}\bracelu\@bLfil\bracerd\braceld\@bRfil\braceru$}
\def\downbracefill{$\m@th\setbox\z@\hbox{$\braceld$}\braceld\@bLfil\braceru\bracelu\@bRfil\bracerd$}
\makeatother

\usepackage{tikz,pgf}
\usetikzlibrary{shapes}
\usetikzlibrary{calc}
\usetikzlibrary{decorations.pathmorphing}
\usetikzlibrary{decorations.pathreplacing,shapes.misc}
\usetikzlibrary{positioning}
\usetikzlibrary{arrows}
\usetikzlibrary{decorations.markings}

\def\be{\begin{equation}}
\def\ee{\end{equation}}
\def\barr{\begin{array}}
\def\earr{\end{array}}

\def\1{\tilde{1}}
\def\2{\tilde{2}}
\def\3{\tilde{3}}

\newcommand{\ba}{\begin{equation}\begin{aligned}}
\newcommand{\ea}{\end{aligned}\end{equation}}

\newcommand{\bml}{\begin{multline}}
\newcommand{\eml}{\end{multline}}

\newcommand{\CC}{\mathbb{C}}

\newcommand{\ZZ}{\mathbb{Z}}
\newcommand{\PP}{\mathbb{P}}

\newcommand{\dd}{\mathrm{d}}
\newcommand{\pd}{\partial}

\newcommand{\Res}{\mathrm{Res}}

\newcommand{\MM}{\mathscr{M}}
\newcommand{\aA}{\mathscr{A}}

\mathtoolsset{showonlyrefs}

\begin{document}

\title{Special geometry on the moduli space for the two-moduli non-Fermat Calabi--Yau}

\author{Konstantin Aleshkin $^{1,2}$}
\author{Alexander Belavin $^{1,3,4}$}

\affil{$^1$ L.D. Landau Institute for Theoretical Physics\\
 Akademika Semenova av. 1-A\\ Chernogolovka, 142432  Moscow region, Russia}

\affil{$^2$ International School of Advanced Studies (SISSA),
 via Bonomea 265, 34136 Trieste, Italy}
\affil{$^3$ Department of Quantum Physics\\
Institute for Information Transmission Problems\\
 Bolshoy Karetny per. 19, 127994 Moscow, Russia}
\affil{$^4$ Moscow Institute of Physics and Technology\\
Dolgoprudnyi, 141700 Moscow region, Russia}

\maketitle
\abstract{ We clarify the recently proposed method for computing a special K\"ahler metric on a Calabi--Yau
complex structure moduli space using the fact that the moduli space is a subspace of a particular Frobenius manifold. We use this method to compute a previously unknown special K\"ahler metric in
a two-moduli non-Fermat model.
}

\flushbottom




\section{Introduction}

To find the low-energy Lagrangian of string theory compactified on a Calabi--Yau (CY)
manifold $X$, one must know the special K\"ahler geometry on the moduli space
of the CY manifold~\cite{Distances, Rolling, S, CO}.
An alternative method for computing the K\"ahler potential in the case
where the CY manifold is given by a hypersurface $W_0(x) = 0$ in a weighted projective space
was recently presented in~\cite{AKBA}.
In this paper, we briefly review this method and use it to compute
the special K\"ahler geometry on the moduli space of a two-moduli non-Fermat threefold considered
in \cite {BCOFHJQ}.
This approach is based on the fact that the moduli space of a CY manifold is a subspace in the particular Frobenius manifold (FM)~\cite{Dub}
that arises from the deformation space of the singularity defined by the LG superpotential $W_0(x) $.
In this paper, we discuss this FM and its connection with the geometry of the CY
moduli space in more detail.

It is known, that an isolated singularity $W_0(x)$ defines an
object called a Milnor ring $R_0$,
Endowed with an invariant pairing this ring becomes a Frobenius algebra.
 Its elements define versal deformations of the singularity.
It is well known that the space of deformations admits a FM structure.
In our case, $W_0(x)$ is a quasihomogeneous polynomial, which defines
a CY manifold in the weighted projective space. For our purposes of computation of the 
CY moduli space geometry we actually need
the Frobenius subalgebra $R_0^Q$ of $R_0$, that is multiplicatively generated only
 by polynomials of the same degree $d$ as $W_0(x)$. These polynomials are related to the complex
 structure deformations.
 The subalgebra $R_0^Q$ consists of elements of degrees $0$, $d$, $2d$ and $3d$, and each
homogeneous component of this algebra is naturally a subgroup of the respective middle Hodge
cohomology $H^{3,0}(X)$, $H^{2,1}(X)$, $H^{1,2}(X)$ and $H^{0,3}(X)$. In our examples,
$R^Q_0$ is actually isomorphic to $H^3(X)$. The FM in which we are interested is precisely the submanifold
of the FM on the versal deformations of $W_0(X)$ restricted to the deformations by the elements
of $R_0^Q$.

The relation to the FM structure allows obtaining the following explicit expression~\cite{AKBA} for the K\"ahler potential in terms of the holomorphic FM metric $ \eta_{\mu\nu}$ in the
flat coordinates and periods of the holomorphic 3-form $\Omega$:
\be \label{maineq}
e^{-K} = \sigma^+_{\mu}(\phi) \eta^{\mu\rho} M^{\nu}_{\rho} \overline{\sigma^-_{\nu}(\phi)}, \quad
\; M = T^{-1} \bar{T}.
\ee
Here, $\phi$ is a parameter on the moduli space, and $\sigma^{\pm}_{ \nu}(\phi)$ are periods of the holomorphic 3-form $\Omega$, defined below
as the integrals of $\Omega$ over some special basis of cycles $\gamma_{\mu}$ in the homology group $ H_3(X)$, which are defined in terms of the connection with the FM.

The constant nondegenerate complex matrix $T$ connects the basis $\sigma^{\pm}_{ \nu}(\phi)$ with another basis $\omega^{\pm}_{ \mu}(\phi)$ defined in~\cite {COGP, BCOFHJQ} as integrals of $\Omega$ over another special basis of the cycles $q_{\nu}$ by the linear relation
\be \label{relT}
\omega^{\pm}_{\mu}(\phi) = T^{\nu}_{\mu} \; \sigma^{\pm}_{ \nu}(\phi).
\ee
The explicit expression for the K\"ahler potential is obtained as follows.
The basis $\omega_{\mu}$ is defined as integrals over the homology
cycles $q_{\mu}$ with real coefficients.
These cycles have a well-defined intersection matrix $C_{\mu\nu}= q_{\mu}\cap q_{\nu}$. The matrix
$M$ and formula~\eqref{maineq} are independent of the specific choice of $\omega_{\mu}$.
The K\"ahler potential $K(\phi)$ on the moduli space with the coordinates $\phi$ and the
holomorphic FM
metric $ \eta_{\alpha\beta}$ can be expressed in the terms of $C_{\mu\nu}$,
the periods $\omega_{\mu}$, and some additional periods
$\omega_{\alpha\mu}(\phi)$ (also defined below) as
\be \label{K}
e^{-K(\phi)} = \omega_{\mu}(\phi) C^{\mu\nu} \overline{\omega}_{\nu}(\phi) \; , \\
\ee
and
\be \label{h}
\eta_{\alpha\beta} = \omega_{\alpha\mu}(0) C^{\mu\nu} \omega_{\beta\nu}(0) \; . \\
\ee
Relation~\eqref{K} was proved in~\cite{Distances, Rolling, S, CO} (see~\cite{CV, Chiodo}
and also~\cite{AKBA} for the proof of relation~\eqref{h}).
Using these two formulas together with relation~\eqref{relT} between the periods
$\omega^{\pm}_{ \mu}(\phi)$ and $\sigma^{\pm}_{ \nu}(\phi)$,
we obtain formula~\eqref{maineq} for the K\"ahler potential.\\

In the beginning of the paper, we briefly review our method~\cite{AKBA} for computing
the moduli space geometry and clarify the role of the invariant Milnor ring $R^Q_0,$ which was
not explicit in~\cite{AKBA}. Then we demonstrate its efficiency by computing the complex structure
manifold geometry for a one-parameter family of quintic threefolds, which was previously found
in~\cite {COGP} by an different approach. For the reader's convenience,
we present our construction for this simple example, which we studied with our method in~\cite{AKBA}.
Our intention is to illustrate the relation between the Milnor ring, the invariant Milnor ring, and
the cohomology for the quintic and its quotient and to show its importance for a proper understanding of the method.

We then use our method to compute a new case of the complex structure geometry
for a two-parameter family of CY threefolds of a non-Fermat type
considered in~\cite {BCOFHJQ}.

\section{Special geometry}

We recall the basic facts about the special K\"ahler geometry and how it arises
on the moduli space of complex structures on CY manifolds (see~\cite{Distances, Rolling, S, CO}).

Let $\MM$ be a K\"ahler $n$-dimensional manifold and $z^1 \cdots z^{n+1}$ be a set of holomorphic
projective coordinates on it.
Then $\MM$ is called a special K\"ahler manifold with special coordinates $z^i$ if
there exists a holomorphic homogeneous function $F(z)$ of degree two in $z$, called a
prepotential, such that the K\"ahler potential $K(z)$ of the moduli space metric is given by
\be e^{-K(z)} = z^a \cdot \frac{\pd \bar{F}}{ \pd \bar{z}^{\bar{a}}} -
\bar{z}^{\bar{a}} \cdot \frac{\pd F}{ \pd z^{a}}.
\ee
Let $X$ be a CY threefold with complex coordinates $y^{\mu}$ ($ \mu=1,2,3$)
and the holomorphic 3-form $\Omega$.
The moduli space of complex structures is
the space of perturbations of the metric on $X$ that preserve the Ricci flatness of the metric.
These deformations are related to the harmonic forms $\chi_{a}\in H^{2,1}(X)$ as
\be
\delta_a g_{\bar{\alpha}\bar{\beta}} \to\chi_{a, \mu\nu\bar\beta}
\sim\Omega_{\mu\nu\lambda}g^{\lambda \bar{\alpha}} \delta_a g_{\bar{\alpha}\bar{\beta}}.
\ee

The Weil--Peterson metric on the CY moduli space is
\be
G_{a\bar{b}} = \int_X \dd^6 y \; g^{1/2} \; g^{\mu \bar{\sigma}} g^{\nu \bar{\rho}}
\delta_a g_{\mu\nu} \delta_{\bar{b}} g_{\bar{\sigma} \bar{\rho}} \;.
\ee
For such CY metric deformations, we obtain~\cite{Distances, Rolling, S, CO}
\be
G_{a\bar{b}} = \frac{\int_X \chi_a \wedge \bar{\chi}_{\bar{b}}}{ \int_X \Omega \wedge \bar{\Omega}}.
\ee
Here, $a$ and $\bar{b}$ are indices of complex coordinates in the moduli space of the CY complex structure.

It follows from the Kodaira lemma~\cite{CO},
\be
\pd_a \Omega = k_a \Omega + \chi_a,
\ee
that this is a K\"ahler metric:
\be
G_{a\bar{b}} = -\pd_a \pd_{\bar{b}} \ln \int_X \Omega \wedge \bar{\Omega}.
\ee
It is convenient to use the basis of periods that are integrals over the Poincar\'e dual symplectic basis
$A^a,B_b\in H_3(X, \ZZ)$:
\ba
A^a \cap B_b = \delta^{a}_{b},\qquad A^a \cap A^b =0,\qquad B_a \cap B_b = 0.
\ea
By defining periods in the symplectic basis as
\be
z^a = \int_{A^a} \Omega, \;\;\; F_b = \int_{B_b} \Omega,
\ee
we obtain
\be
e^{-K} = \int_X \Omega\wedge \bar{\Omega} =
z^a \cdot \bar{F}_{\bar{a}} - \bar{z}^{\bar{a}} \cdot F_{ a}.
\ee

It also follows from the Kodaira lemma that
$ F_a(z) = \frac{1}{2} \pd_a F(z),$ where $F_a(z) = \frac12 z^a F_a(z)$.
Therefore,
\be e^{-K(z)} = z^a \cdot \frac{\pd \bar{F}}{ \pd \bar{z}^{\bar{a}}} -
\bar{z}^{\bar{a}} \cdot \frac{\pd F}{ \pd z^{a}}.
\ee
Hence, $G_{a\bar{b}}$ is the special K\"ahler metric with the prepotential $F(z)$.

Using the notation $\Pi = \begin{pmatrix} F_{\alpha}, z^{b} \end{pmatrix}$ for the vector of periods,
we write the expression for the K\"ahler potential as
\be \label{symplKahler}
e^{-K(z)} = \Pi_a \Sigma^{ab} \bar{\Pi}_b,
\ee
where the symplectic unit $\Sigma$ is an inverse intersection matrix for the cycles $(A^a, B_a)$.
We can therefore transform this expression to form~\eqref{K}:
\be
e^{-K(\phi)} = \omega_{\mu}(\phi) C^{\mu\nu} \bar{\omega}_{\nu}(\phi),
\ee
where $\omega_{\mu}(\phi)$ are the integrals of $\Omega$ over the
arbitrary homology basis $q^{\mu}$.

\section{The CY manifold as a hypersurface in a weighted projective space}
In what follows, we concentrate on the case where the CY manifold is realized
as a hypersurface in a weighted projective space or its quotient by some discrete group.
Let $x_1, \ldots , x_5$ be homogeneous coordinates in a weighted projective space
$\PP^{4}_{(k_1, \ldots, k_5)}$ and
\be \label{X}
X = \{ (x_1: \ldots: x_5) \; \in \PP^{4}_{(k_1, \ldots, k_5)} | \; W_0(x) = 0 \}.
\ee
Also let $W_0(x)$ be a quasihomogeneous polynomial defining an isolated singularity at the origin,
\be
W_0(\lambda^{k_i} x_i) = \lambda^d W_0(x_i).
\ee
Then $X$ is a CY manifold if
\be
\operatorname{deg} W_0(x) = d = \sum_{i=1}^5 k_i.
\ee

The moduli space of complex structures is then given by (a quotient of) the space of
homogeneous polynomial
deformations of this singularity modulo coordinate transformations:
\be
W(x, \phi) = W_0(x) + \phi_0 \prod x_i + \sum_{s=1}^{N-1} \phi^s e_s(x),
\ee
where $N$ is a number of polynomial deformations and
 $e_s(x)$ are polynomials of the same weight (degree) as $W_0(x)$. They are invariant
under the $\ZZ_d$ action $\alpha \cdot e_{\mu}(x_i) = e_{\mu}(\alpha^{k_i} x_i)$,
$\alpha^d = 1.$

The holomorphic volume form $\Omega$ can be written~\cite{COGP, BCOFHJQ} as a residue of a 5-form in the underlying
affine space $\CC^5$:
\begin{align}\label{omega}
\Omega = \frac{x_5 \dd x_1\wedge \dd x_2\wedge \dd x_{3}}{\pd W(x)/\pd x_{4}} &=
\Res_{ W(x) = 0} \frac{x_5 \dd x_1\cdots\dd x_{4}}{W(x)}
\nonumber
\\
&=\frac{1}{2\pi i}
\oint_{|x_5|=\delta} \Res_{ W(x) = 0} \frac{\dd x_1\cdots\dd x_{5}}{W(x)}\;.
\end{align}
Taking this explicit expression for $\Omega$, we obtain a basis
of periods $\omega_{\mu}(\phi)$ as follows~\cite{COGP, BCOFHJQ}.
We take a so-called fundamental cycle $q_1$, which can be described as a torus in
the large complex structure limit (LCSL)
$\phi_0 \gg 1$ (for simplicity, the other parameters $\phi^s=0$):
\be
W(x,\phi) = W_0(x) + \phi_0 \prod x_i .
\ee

In the same limit, $Q_1 := \{|x_i| = \delta_i\} \in H_5(\CC^5 \backslash W(x) = 0)$ is a 5-dimensional
torus surrounding
the hypersurface $W(x) = 0$ in $\CC^5$. It is chosen to represent the integral over a three-dimensional torus $q_1 \in
H_3(X)$. The fundamental period is then
\be
\omega_1(\phi) := \int_{q_1} \Omega = \int_{Q_1} \frac{\dd x^1\cdots\dd x^5}{W(x,\phi)}
\ee
and is given by a residue as a series in $1/\phi_0$.

More periods $\omega_{\mu}$ can be obtained
as analytic continuations of $\omega_1$ in $\phi$.
This can be done by continuing $\omega_1(\phi)$ in a small $\phi_0$
region using Barnes' trick and subsequently using the symmetry of $W_0(x)$~\cite{BCOFHJQ}.

There is a group of \textit{phase} symmetries $\Pi_{X}$ defined as a group acting
diagonally on $x_i$ and preserving $W_0(x)$.
When $W_0(x)$ is deformed, this group acts on the parameter space
with an action $\aA$ such that
$$
W(g\cdot x,~ \aA(g) \cdot \phi_0) = W(x, \phi).
$$
The moduli space is then at most a quotient $\{\phi^s\}_{0\le s\le N-1}/\aA$ of the parameter space.
We can thus define a set of periods by analytic continuation,
\be
\omega_{\mu_g}(\phi) = \omega_{1}(\aA(g) \cdot \phi_0), \quad g \in G_X.
\ee
In examples, this construction gives all the periods of a volume form of $X$.
We can also consider quotients $X/H$ where $H \subset \Pi_X$ is an admissible subgroup,
which is a subgroup that preserves the volume form $\Omega$. In this case, we should consider
only $H$-invariant deformations and corresponding quantities.

It is important that we can represent the periods $\int_{q_{\mu}} \Omega$
as integrals of the complex oscillatory form. Starting from
\be
\omega_{\mu}(\phi) := \int_{q_{\mu}}\Omega = \int_{Q_{\mu}} \frac{\dd^5 x}{W(x)},
\ee
we can rewrite them as
\be \label{osc}
\int_{Q_{\mu}} \frac{\dd^5 x}{W(x)}
= \int_{Q^{\pm}_{\mu}} e^{\mp W(x)} \dd^5 x ,
\ee
where
$ Q_{\mu}^{\pm} \in H_5(\CC^5, \; \mathrm{Re}\,W_0(x) = \pm\infty)$~\cite{AVG, Vas, Kon}.
Indeed, both sides of the equation satisfy the same differential equations which can be
obtained from computations in the Milnor ring. Moreover, there is the isomorphism~\cite{Vas}
\be
H_{3}(X) \to H_5(\CC^5 \backslash W(x) = 0) =
H_5(\CC^5, \; \mathrm{Re}\,W_0(x) = \pm \infty)_{w\in d\cdot\ZZ} \;.
\ee
Therefore, more precisely, we have $Q_{\mu}^{\pm} \in H_5(\CC^5, \; \mathrm{Re}\,W_0(x)=\pm \infty)_{w\in d\cdot\ZZ}$,
which is a subgroup of $ H_5(\CC^5, \; \mathrm{Re}\,W_0(x) = \pm \infty) $ defined below.

\section{Frobenius manifold}

It is known that compactification of the superstring theory on a CY manifold is deeply
connected with the compactification on the $N{=}2$ superconformal theories~\cite{Gep}.
When the latter are Landau--Ginzburg theories, they are connected with singularity
theory~\cite{LVW, Mart, Vafa}.

Using this connection, we can extract the information about the special geometry of the CY moduli space
from the Milnor ring of the defining singularity $W_0(X)$~\cite{AVG,Blok}:
\be
R_0 = \frac{\CC[x_1, \ldots, x_5]}{\pd_1 W_0(x) \cdot \ldots \cdot \pd_5 W_0(x)}.
\ee
In fact, we need to consider not the whole Milnor ring but its $Q:=Z_d$-invariant subring $R_0^Q$.
This subring is multiplicatively generated by marginal deformations of the singularity,
i.e., by those deformations that have the same weight $d$ as $W_0(x)$ and
correspond to the complex structure moduli of CY. This subring consists of the elements of the
Milnor ring $R_0$ whose weights are integer multiples of $d,$ that is $0,d,2d,3d$.
 We note that in many cases,
the dimension of the subring $R_0^Q$ is equal to the dimension of the
homology group $H_3(X)$, where $X$ is the CY defined using the polynomial
$W_0(x)$ in~\eqref{X}. But in general, $\dim R^Q_0 \le \dim H^3(X).$ In the case of
the strict inequality, we restrict our attention to the subspace of $H^3(X)$ that is isomorphic
to $R^Q_0$ without noting this explicitly.

We let $e_{s}(x)$ (with Latin indices ${s}$) denote the elements that
correspond to the complex structure deformations of $X$ and $e_{\mu}(x)$ (with
Greek indices ${\mu}$) denote all elements of the basis of $R_0^Q$.
In the example where $X$ is a quintic threefold, we have the dimension
$\dim R_0= 1024$ of the whole Milnor ring, the dimension $\dim R_0^Q=\dim H_3(X)=204$
of the Q-invariant subring, and the dimension $\dim M_c=101$ of the subspace of
marginal deformations.

There exists a natural multiplication with structure constants $C_{\mu\nu}^{\sigma}$ in $R_0^Q$ and a pairing $\eta_{\mu\nu}$ that makes $R_0^Q$ a Frobenius algebra:
\ba
\eta_{\mu\nu} = \Res \frac{e_\mu \cdot e_\nu}{\pd_1 W_0(x) \cdots \pd_5 W_0(x)}, \\
C_{\mu\nu\lambda} = C_{\mu\nu}^{\sigma} \eta_{\sigma\lambda} = \Res \frac{e_\mu \cdot e_\nu \cdot e_\lambda}{\pd_1 W_0(x) \cdots \pd_5 W_0(x)}.
\ea
We consider the space of deformations of the singularity
\be
W(x) = W_0(x) + \sum t^{\mu} e_{\mu}(x),
\ee
where $e_{\mu}(x)$ belongs to the Q-invariant subring $R^Q_0$ of the Milnor ring.
The structure of the FM $M_F$~\cite{Dub}
with multiplication structure constants $C_{\mu\nu}^{\rho}(t)$ in the
ring $R^Q$ defined by the deformed singularity $W(x)$ arises on the space with the parameters $t^{\mu}$:
\be
R = \frac{\CC[x_1, \ldots, x_5]}{\pd_1 W(x) \cdot \ldots \cdot \pd_5 W(x)} \;, \quad R^Q \text{ is a }\ZZ_d
\text{ invariant subset of } R \; .
\ee
It has a Riemanian flat metric $h_{\mu\nu}(t)$ and the metric $h_{\mu\nu}(t=0)$ equal to $\eta_{\mu\nu}$.
The structure constants are the third derivatives of the Frobenius potential $F(t)$.
This Frobenius potential coincides with the holomorphic prepotential of the special
K\"ahler geometry when restricted to the marginal subspace.
They are naturally the Yukawa couplings of the corresponding fields~\cite{CO}. We note
that the FM in question is naturally a complex manifold and the metric $h_{\mu\nu}(t)$ and
structure constants $C^{\rho}_{\mu\nu}(t)$ are holomorphic.

We note that the small FM $M_F$ is a
submanifold of the total FM arising from the whole Milnor ring $R_0$. The manifold $M_F$
is connected with the cohomology of $X$ and is used in our approach.

Only quasihomogeneous deformations $W(x)$ define a hypersurface in a weighted projective space.
The marginal deformations $W_0(x) + \sum \phi^{s} e_{s}(x)$ define a subspace of
the FM connected with $ W_0$.
This subspace $ M_c$ of the FM coincides with the moduli space of the CY manifold (at least locally and maybe after some orbifolding).
This fact is very important for computing the special geometry: it allows expressing the matrix $C_{\mu\nu}$ in terms of the FM metric $\eta_{\mu\nu}$~\cite{AKBA}.

\section{ The idea for computing the periods}
We can now relate the oscillatory form of the period integrals~\eqref{osc} to
the FM structure and to the holomorphic metric $\eta_{\mu\nu}$.

We consider the differentials $D^+$ and $D^-$ given by
\be
D^{\pm} = D^{\pm}_{W_0} = \dd \pm \dd W_0\wedge.
\ee
They define cohomology subgroups $H^5_{D^{\pm}}(\CC^5)_{w\in d\cdot\ZZ}$ on the space of
differential forms of the weight $d\cdot \ZZ$. These subgroups are isomorphic
as linear spaces to the ring $R_0^Q$
$$
e_{\mu}(x) \to e_{\mu}(x) \dd^5 x.
$$
Moreover, $R_0^Q$ acts naturally on $H^5_{D^{\pm}}(\CC^5)_{w\in d\cdot\ZZ}$ by multiplication
(in terms of representatives).
The cohomology subgroups $H^5_{D^{\pm}}(\CC^5)_{w\in d\cdot\ZZ}$ are dual to the
homology subgroups $H_5(\CC^5, \mathrm{Re}\,W_0(x) = \mp \infty)_{w\in d\cdot\ZZ}$
that consist of cycles
$\Gamma^{\pm}_{\mu} $ $\subset H_5(\CC^5, \; \mathrm{Re}\,W_0(x) = \pm \infty)$ with a nondegenerate
pairing with $e_{\nu}(x) \dd^5 x \in H^5_{D^{\pm}}(\CC^5)_{w\in d\cdot\ZZ}$ defined as
\be
\langle \Gamma^{\pm}_{\mu}, \; e_{\nu}\dd^5 x \rangle=
\int_{\Gamma^{\pm}_{\mu}} e_{\nu} \cdot e^{\mp W_0(x)} \dd^5 x .
\ee

The $H^5_{D^{\pm}}(\CC^5, \; \mathrm{Re}\,W_0(x)=\pm \infty)_{w\in d\cdot\ZZ}$ are invariant under $x_i \to e^{2\pi i k_i/d}x_i$.
We have a group isomorphism $H^5_{D^{\pm}}(\CC^5, \; \mathrm{Re}\,W_0(x)=\pm \infty)_{w\in d\cdot\ZZ}= H_3(X)$ and therefore $R^Q \simeq H^3(X)$.
This isomorphism maps the weight filtration in the left-hand side to the Hodge filtration in the right-hand side,
i.e., cycles of weight $k\cdot d$ correspond to differential forms in $H^{3-d,d}(X)$.

A possible choice of cycles $\Gamma^{\pm}_{\mu}$ is
\be
\int_{\Gamma^{\pm}_{\mu}} e_{\nu} \cdot e^{\mp W_0(x)} \dd^5 x = \delta^{\mu}_{\nu}.
\ee
A convenient computation technique can be used to find the periods represented as the oscillatory integrals
\be
\int_{\Gamma^{\pm}_{\mu}} e_{\nu} \cdot e^{\mp W(x,\phi)} \dd^5 x ,
\ee
where $W(x, \phi) = W_0(x) + \sum_{s=0}^{\mu} \phi^s e_s(x)$. Expanding the integrand
in a series in $\phi^s$, we obtain integrals of the type
$ \int_{\Gamma^{\pm}_{\mu}} P(x) e^{-W_0(x)} \dd^5 x$. Here, $P(x)$ is a product of $e_s(x)$ and
is hence a Q-invariant monomial.

The technique for computing such integrals, previously used to compute the flat coordinates in the topological CFT~\cite {BDM, BU, Kon, BB}, is based on the fact that
\be
\int_{\Gamma^{\pm}_{\mu}} P(x) e^{\mp W_0(x)} \dd^5 x =
\int_{\Gamma^{\pm}_{\mu}} \tilde{P}(x) e^{\mp W_0(x)} \dd^5 x
\ee
if the forms in the integrands are equivalent in the $D^{\pm}$ cohomology,
\be
P(x)\dd^5 x -\tilde{P}(x)\dd^5 x = D^{\pm} U.
\ee
Using this, we can easily see that an arbitrary Q-invariant form $P(x) \dd^5 x$ is
reducible to a linear combination of $e_{\mu}(x)\dd^5 x$, where the $e_{\mu}(x)$ form the basis
of $R^Q_0$.

Computing integrals of the type
$ \int_{\Gamma^{\pm}_{\mu}} P(x) e^{-W_0(x)} \dd^5 x$
thus becomes a linear problem of expanding $ P(x)\dd^5 x$ over the basis of
$H^5_{D^{\pm}}(\CC^5)_{w\in d\cdot\ZZ}$.

\section{\texorpdfstring{Finding $C^{\mu\nu}$ and the K\"ahler potential}
{Finding \$C\000\136\{\000\134mu\000\134nu\}\$ and the K\"ahler potential}}

We use the relation of the CY moduli space to the FM structure
to find the intersection matrix of the cycles $C_{\mu\nu}=
q_{\mu} \cap q_{\nu}=Q_{\mu}^+ \cap Q_{\nu}^- $.
For this, we introduce a few new sets of periods $ \omega^{\pm}_{\alpha, \mu}(\phi) $
as integrals of
$e_{\alpha}(x) \dd^5 x$ $\in H^5_{D^{\pm}}(\CC^5, \; \mathrm{Re}\,W_0(x)=\pm \infty)_{w\in d\cdot\ZZ}$
over the cycles $Q^{\pm}_{\mu} \in H_5(\CC^5, \; \mathrm{Re}\,W_0(x) = \pm\infty)_{w\in d\cdot\ZZ} $
defined above:
\be
\omega^{\pm}_{\alpha\mu}(\phi) = \int_{Q^{\pm}_{\mu}} e_\alpha(x) \, e^{\mp W(x, \phi)} \dd^5 x.
\ee

The periods $\omega^{\pm}_{1\mu}(\phi)$ coincide with the periods $\omega^{\pm}_{\mu}(\phi)$ because we assume that $e_1(x)=1$ denotes the unity in the ring $R$.

The additional periods allow computing $C^{\mu\nu}$ because of its relation
to the FM metric $\eta_{\alpha\beta}$~\cite{CV, Chiodo}:
\be
\eta_{\alpha\beta} =\omega^+_{\alpha, \mu}(t=0) C^{\mu\nu }\omega^-_{\beta, \nu}(t=0).
\ee

From this formula, we can obtain the expression for $C^{\mu\nu}$
if we know the values of $\omega^+_{\alpha, \mu}(t=0)$ for all $ \alpha$.
As follows from the definition of $\omega^{\pm}_{\alpha\mu}(\phi)$,
we can express $\omega^+_{\alpha, \mu}(\phi=0)$ in terms of
the derivatives of the periods $\omega^{\pm}_{\mu}(\phi)$ with respect to $\phi$ up to the third order
at $\phi=0$ because the basis elements $e_{\alpha}(x)$ of the invariant subring $R_0^Q$
of the Milnor ring can
be chosen as products of the marginal deformations $e_s(x)$, which are related by
\be
e_s(x) \cdot e^{W(x)} \dd^5 x = \frac{\pd}{\pd \phi^s} \; e^{W(x)} \dd^5 x \; .
\ee

Setting $\omega^{\pm}_{\alpha,\mu}(\phi=0):=(T^{\pm})^{\alpha}_{\mu}$,
we rewrite the above relation as
\be \label{etac}
\eta^{\mu\nu}= (T^+)_{\rho}^{\mu} \; C^{\rho\sigma} \; (T^-)^{\nu}_{\sigma},
\ee
which gives the expression for the intersection matrix $C^{\rho\sigma}$ in terms of
$\eta^{\mu\nu}$ and the matrix $T$. The result can be substituted in the K\"ahler potential formula
$$e^{-K(\phi)} = \omega_{\mu}(\phi) C^{\mu\nu} \bar{\omega}_{\nu}(\phi)$$
to obtain the explicit expression for $K(\phi)$.

To obtain a more convenient expression for $K(\phi)$, we define one more
basis of periods $\sigma^{\pm}_{\mu}(\phi)$ as integrals over the cycles
$\Gamma^{\pm}_{\mu} \in H_5(\CC^5, \; \mathrm{Re}\,W_0(x) = \pm\infty)_{w\in d\cdot\ZZ} $
defined above:
\be \label{sigmagen}
\sigma^{\pm}_{\mu}(\phi) = \int_{\Gamma^{\pm}_{\mu}} e^{\mp W(x,\phi)} \dd^5 x .
\ee
This basis comprises eigenvectors of the phase symmetry action $\aA \, : \,
\Pi_X \times \{\phi^s \}\to\{\phi^s\}$.
Having the oscillatory representation for the periods $\sigma^{\pm}_{\mu}(\phi)$
over the corresponding cycles $\Gamma^{\pm}_{\mu}$, we can define additional integrals
$\sigma^{\pm}_{\alpha, \mu}(\phi)$ over the same cycles as
\ba \label{sig}
\sigma^{\pm}_{\alpha, \mu}(\phi) = \int_{\Gamma^{\pm}_{\mu}} e_{\alpha}(x) \, e^{\mp W(x,\phi)} \dd^5 x.
\ea
It follows from $e_1(x) = 1$ that $\sigma^{\pm}_{1\mu} = \sigma^{\pm}_{\mu}$.
Because of our choice of the cycles $\Gamma^{\pm}_{\mu}$, we also have
$\sigma^{\pm}_{\alpha, \mu}(t=0) = \delta_{\alpha, \mu}$.

Because $\omega^{\pm}_{ \mu}(\phi)$ and $\sigma^{\pm}_{ \nu}(\phi)$ are
both bases of periods defined as integrals over cycles in $H_5(\CC^5, \; \mathrm{Re}\,W_0(x) = \pm\infty)_{w\in d\cdot\ZZ} $,
they are related by some constant matrix. Moreover, this matrix coincides with the matrix $ (T^{\pm})_\alpha^{\nu} = \omega^{\pm}_{\alpha\mu}(\phi=0)$,
\be
\omega^{\pm}_{\mu}(\phi) = (T^{\pm})^{\nu}_{\mu} \; \sigma^{\pm}_{ \nu}(\phi).
\ee
The same relation holds for the other integrals $\omega^{\pm}_{\alpha \mu}(\phi)$ and $\sigma^{\pm}_{\alpha \nu}(\phi)$ over the same cycles
for each $\alpha$,
\be
\omega^{\pm}_{\alpha\mu}(\phi) = (T^{\pm})^{\nu}_{\mu} \; \sigma^{\pm}_{\alpha\nu}(\phi)\;.
\ee
The last formula together with $\sigma^{\pm}_{\alpha\mu}(0) = \delta_{\alpha\mu}$ (which
is by definition~\eqref{sig}) implies that $(T^{\pm})_\alpha^{\nu} = \omega^{\pm}_{\alpha\mu}(\phi=0)$.

Having computed the matrix $T^{\mu}_{\mu}$, we use~\eqref{etac} to express
the intersection matrix $C^{\rho\sigma}$ in terms of this matrix and the known Frobenius metric
$\eta^{\mu\nu}$.

We thus obtain the main statement that
\be \label{kphi}
e^{-K(\phi)} = \sigma_\mu(\phi) \;\eta^{\mu\nu} \; M_{\nu}^\lambda \; \overline{\sigma^-_{\lambda}
(\phi)},
\ee
where the matrix $M^a_b = (T^{-1})^a_c \; \bar{T}^c_b$.
This expresses the K\"ahler potential $K$ explicitly in terms of the periods $\sigma_\mu(\phi)$,
the FM metric $\eta_{\mu\nu}$, and the matrix $T^\mu _\nu$.
All these data can be computed exactly, as explained above.

We also note two points. First, we can also find a symplectic basis of cycles
by applying the Gram--Schmidt process to the obtained intersection matrix. Second,
formula~\eqref{kphi} can be used without explicitly computing $T$ if
the real structure matrix $M$ can be found from some other argument.

\section{Example: Quintic threefold}

We consider the one-parameter family of CY manifolds defined as
\begin{equation} \label{dpencil}
X_{\psi} = \{x_i \in \PP^4 \; | \; W_{\psi}(x) =
x_1^5+x_2^5+x_3^5+x_4^5+x_5^5
-5\psi x_1x_2x_3x_4x_5 = 0 \} \; ,
\end{equation}
which was considered in detail in~\cite{COGP}.
The phase symmetry in this case is $\Pi_X = \ZZ_5^5$. The full Milnor ring $R_0$ is 1024-dimensional
and consists of all polynomials in five variables where each of them has the degree less than four.

As explained above, we need not the whole Milnor ring but only its $\ZZ_5$-invariant subring,
which has the dimension $\dim R^Q_0 = 204$ and is isomorphic to $H_3(X)$.
It comprises the polynomials in $R_0$ of degrees 0, 5, 10, and 15. The
101 fifth-degree polynomials (marginal deformations) correspond to the complex structure moduli.

To build the one-dimensional family~\eqref{dpencil}, we take a subgroup
$H = \ZZ^3_5 \subset \Pi_X$ of phase symmetries and consider $\hat{X}=X/H$, which turns out to be
a mirror of $X$. Family~\eqref{dpencil} is the maximal deformation surviving this
factorization. The induced action $\aA$ of the phase symmetry group on
the one-dimensional space $\{\psi \}$ is $\ZZ_5: \psi \to e^{2\pi i/5} \psi$.
The invariant Milnor ring is four-dimensional:
$\hat{R}^Q_0 = \langle 1, \; \prod_i x_i, \; \prod_i x^2_i, \; \prod_i x^3_i \rangle$.

Having the invariant Milnor ring, we define the corresponding cohomology group
and dual cycles $\Gamma^{\pm}_{\mu}$.
Using the recursion procedure for~\eqref{sigmagen}, we obtain
\begin{equation}
\sigma^{\pm}_{{\mu}}(\psi) = \frac{(\pm 1)^{{\mu}-1}}{\Gamma(\mu/5)^5 5^{\mu}\psi}\sum_{n=0}^{\infty}
\frac{\Gamma^5(n+\mu/5)} {\Gamma(5n+\mu)}(5\psi)^{5n+\mu}
=\frac{(\pm \psi)^{\mu-1}}{\Gamma(\mu)} + O(\psi^{\mu+3}).
\end{equation}

The fundamental period for the quintic is a residue of a holomorphic 3-form $\Omega$,
\be
\frac{x_5 \dd x_1\wedge\dd x_2\wedge\dd x_3}{\pd P_{\psi}/\pd x_4},
\ee
defined as an integral over a cycle $q_1$.
Its analytic continuations give the whole basis of periods
in terms of integrals over a basis of cycles $\in H_3(\hat X)$:
\be
\omega_{\mu}(\psi) =
\sum_{m=1}^{\infty}
\frac{e^{4\pi i m/5}\Gamma(m/5)(5e^{2\pi i (\mu-1)/5}\psi)^{m-1}}{\Gamma(m)\Gamma^4(1-m/5)},
\quad |\psi|<1.
\ee
The matrix $T$ is given by
\be
T^{\mu}_{\alpha} =\omega_{\alpha\mu}(0) =
\frac{\pd^{\alpha-1}}{\pd \psi^{\alpha-1}}\omega_{\mu}(0) =
\frac{ 5^{\alpha-1} e^{2\pi i ((\alpha-1) (\mu-1)+2\alpha)/5} \Gamma( \alpha/5) }
{\Gamma^4(1-\alpha/5)}.
\ee
The holomorphic metric $\eta_{\mu\nu}$ on the corresponding FM
is just the coefficient of the element of maximum degree in the decomposition of
$e_{\mu}(x)\cdot e_{\nu}(x)$ in the monomial basis of $\hat{R}^Q_0$. In our case, it is
\be
\eta = \mathrm{antidiag}(1,1,1,1).
\ee

Finally, we obtain $\hat{\eta} = \eta T^{-1}\bar{T}$ and the K\"ahler potential for the metric
\begin{align}
e^{-K(\psi)} ={}& \frac{\Gamma^5(1/5)}{125\Gamma^5(4/5)} \sigma^{+}_{11}\overline{\sigma^{-}_{11}}
+\frac{\Gamma^5(2/5)}{5\Gamma^5(3/5)} \sigma^{+}_{12}\overline{\sigma^{-}_{12}}
\nonumber
\\
&{}+\frac{5\Gamma^5(3/5)}{\Gamma^5(2/5)}\sigma^{+}_{13}\overline{\sigma^{-}_{13}} +
\frac{125\Gamma^5(4/5)}{\Gamma^5(1/5)}\sigma^{+}_{14}\overline{\sigma^{-}_{14}}.
\end{align}
In particular,
\be
G_{\psi \overline{\psi}}(0) = 25\frac{\Gamma^5(4/5)\Gamma^5(2/5)}{\Gamma^5(1/5)\Gamma^5(3/5)},
\ee
which coincides with result in \cite {COGP}.

\section{Two-moduli non-Fermat threefold}

The two-moduli non-Fermat threefold is constructed from the hypersurface
\begin{align} \label{nonfermat}
X = \{&x_i \in \PP^4_{(3,2,2,7,7)} \; |
\nonumber
\\
&W_{\phi}(x) =x_0^7+x_1^7x_3+x_3^3+x_2^7x_4+x_4^3-
\phi_0 x_1x_2x_3x_4x_5 + \phi_1 x_0^3x_1^3x_2^3 = 0 \}.
\end{align}
This example, which was considered in \cite {BCOFHJQ}, is interesting because
it is not of the Fermat type.\footnote{In particular, it is not described
by a product of $N{=}2$ minimal models, and its mirror is obtained from a
seventh-degree hypersurface in a different projective space
$\PP^4_{(1,1,1,2,2)}.$} The weight of the singularity is equal to $d=21$.
The phase symmetry is $\ZZ_{21}^2\times\ZZ^7$. We again consider a quotient
$\hat{X} = X/H$ by the $H=\ZZ_{21}$ action
\be
H := (\ZZ_{21} \; : \; 12,2,0,7,0)
\ee
The Hodge numbers are $h_{1,1}(\hat{X}) = 95$ and $h_{2,1}(\hat{X}) = 2$.
The two--parameter family~\eqref{nonfermat} is the maximum deformation
surviving the factorization. The induced action $\aA$ on
the two-dimensional space $\{\phi_0, \phi_1 \}$ is $\ZZ_7 :\phi_0 \to \alpha \phi_0, \;
\phi_1 \to \alpha^3 \phi_1$, where $\alpha^7 = 1$ is a primitive root.
We note that 7 is a weight of the mirror singularity, as explained in~\cite{BCOFHJQ}.

Analytic continuations of the fundamental period give the full basis of periods
in a basis of cycles with integral coefficients,
\begin{align}
\omega_{\mu}(\psi) =
-\frac17 \sum_{n=1}^{\infty}
e^{6\pi i n/7} \frac{(\alpha^{\mu-1}\phi_0)^{n-1}}{\Gamma(n)}\sum^{\infty}_{m=0}
\frac{e^{-3i\pi m/7} \Gamma\left(\frac{n+3m}{7}\right)}{\Gamma^2\left(1-\frac{n+3m}{7}\right)
\Gamma^2\left(1-\frac{2n-m}{7}\right)} \frac{(\alpha^{3(\mu-1)}\phi_1)^m}{m!},
\\
|\phi_0|, \, |\phi_1| \ll 1.
\nonumber
\end{align}

We now perform the Milnor ring computations to obtain the metric $\eta$. If we set
\be
e_2(x) = x_0x_1x_2x_3x_4, \; e_3(x) = x_0^3x_1^3x_2^3,
\ee
then the $H$-invariant subring of $R^Q_0$ is generated by $e_2$ and $e_3$.
It is easy to compute the relations
\be
e_3^2=0 \qquad e_2^3 = 0.
\ee
Hence, the vector space basis of this subring is
\be
e_1, \; e_2, \; e_3, \;e_4=e_2^2, \;e_5=e_2 e_3, \; e_6 = e_2^2 e_3.
\ee
The last one is of the highest degree 63, and the metric in this basis is therefore
$\eta \simeq \mathrm{antidiag}(1,1,1,1,1,1)$.

Taking the first four terms of the expansion of the above periods, we obtain
\be
T^{\mu}_{\nu} = A(\nu) \alpha^{k_{\nu}(\mu-1)}, \qquad k_{\nu} = (1,2,4,3,5,6),
\ee
and
\be
A(\nu) =
\alpha^{2m_{\nu} - n_{\nu}/2} \frac{(-1)^{n_{\nu-1}}\Gamma\left(\frac{n_{\nu}+3m_{\nu}}{7}\right)}
{\Gamma^2\left(1-\frac{n_{\nu}+3m_{\nu}}{7}\right)
\Gamma^2\left(1-\frac{2n_{\nu}-m_{\nu}}{7}\right) },
\ee
where $(n_{\nu}, m_{\nu})= ((1,0), (2,0), (1,1),
(3,0), (2,1), (3,1))$ correspond to our choice of the basis.

Finally, we obtain $\hat{\eta} = \eta T^{-1}\bar{T}$ and the K\"ahler potential for the metric
\begin{align}
e^{-K(\psi)} ={}& \gamma^3(1/7)\gamma^2(2/7)
\sigma^{+}_{11}\overline{\sigma^{-}_{11}} + \gamma^3(2/7)\gamma^2(4/7)
\sigma^{+}_{12}\overline{\sigma^{-}_{12}}
+ \gamma^3(4/7)\gamma^2(1/7)\sigma^{+}_{13}\overline{\sigma^{-}_{13}}
\nonumber
\\
&{}+\gamma^3(3/7)\gamma^2(6/7)
\sigma^{+}_{14}\overline{\sigma^{-}_{14}} + \gamma^3(5/7)\gamma^2(3/7)
\sigma^{+}_{15}\overline{\sigma^{-}_{15}}
+ \gamma^3(6/7)\gamma^2(5/7)\sigma^{+}_{16}\overline{\sigma^{-}_{16}}\;,
\end{align}
where $\gamma(x)=\Gamma(x)/\Gamma(1-x)$.

In particular, the K\"ahler metric has the form
\be
G(0) =
\begin{pmatrix}
\gamma^3\left(\frac67\right) \gamma^2\left(\frac47\right) \gamma\left(\frac27\right) & 0\\
0 & \gamma^3\left(\frac47\right) \gamma^2\left(\frac57\right) \gamma\left(\frac67\right) \\
\end{pmatrix}.
\ee



\paragraph{ Acknowledgments}
A.~B.~is grateful to F.~Quevedo for the
interesting discussions and hospitality at ICTP, where this work was finished.
The work of A.~B.~on the main results presented in sections 1--5 was
 performed at IITP with the financial support of the Russian Science Foundation
 (Grant No.14-60-00150). The work of K.~A.~was supported by the Foundation for
 the Advancement of Theoretical Physics ``BASIS.''

\printbibliography

\end{document}